\documentclass[12pt]{article}
\usepackage{epsfig}

\catcode`\@=11
\catcode`\@=12

\title{{\small \hfill IMSc-2003/05/08}\\
\textbf{ Quark Propagator and Chiral Symmetry with String Tension}}
\author{ Ramesh Anishetty\footnote{ramesha@imsc.res.in}~ and ~Santosh Kumar Kudtarkar\footnote{sant@imsc.res.in}\\
 The Institute of Mathematical Sciences, Chennai 600 113, India\\}

\date{\today}
\begin{document}
\maketitle
\begin{abstract}
\indent
General properties of the light and heavy quark propagators have been 
investigated in the context of  string tension interaction. Confinement,
 chiral symmetry breaking, spectral properties of the propagator are 
analytically studied and numerically validated. We show that the 
propagator is analytic in the infrared region even for massless 
quarks with a non zero radius of convergence. Emergence of more 
than one mass scale is exemplified. Massless limit of the quark 
propagator does exhibit critical behaviour. 
\end{abstract}

\newpage

\section{Introduction}
Confinement of quarks and their transient behaviour as dictated by 
the quark propagator is being investigated in a theory \cite{ramesha} which shows
 many features of QCD. We  have considered colored quarks in QCD
 with renormalisable string tension in the large $N$ expansion. 
We compute the behaviour of the quark propagator from first 
principles in this quantum field theory as a function of momentum,
 renormalised mass and coupling constant.
 
 The presence of string tension results in the absence of the pole 
in the quark propagator thus manifesting the non-existence of 
quarks as asymptotic states in the quantum field theory. 
In addition for any renormalised mass parameter $m\geq0$,
 we find that the analytic structure of the propagator is non-trivial. 
Firstly for space like momenta ($p^2<0$) the quark propagator is 
analytic. For any renormalised mass there exists a Taylor series 
solution in $p^2$ to the Schwinger-Dyson(SD) equation. This 
series is numerically estimated to converge for $p^2<\tilde{m}^2\neq0$ 
where $\tilde m$ is called the threshold mass. In all, 
associated with a quark, there are three mass parameters namely,
the renormalised mass $m$,  the threshold mass $\tilde m$ and  
the asymptotic mass $M$ which capture the behaviour of the quark
 propagator for space like region. These three masses are very 
distinct and are estimated from phenomenology. We also discuss 
how the so called constituent mass in various bound states is 
related to $\tilde m$. We find for heavy quarks where the 
renormalised mass scale $m^2$ is larger than string tension 
$\sigma$, there is a qualitative difference in the analytic 
properties which can be physically interpreted as, in space-time 
light quarks have zero transient time existence while heavy 
quarks are almost about to be free and hence have relatively 
long time transient existence.
  
 Massless quark theory does exhibit spontaneous chiral symmetry
 breaking in the presence of string tension and the theory is 
self-consistently defined with out the need for ultraviolet(UV) 
renormalisation unlike the massive case. Consequently the quark 
propagator  function is not a continuos function of $m$ at $m=0$.
 The spin independent quark propagator, due to Goldstone phenomenon, 
is also the massless pion wave function with the asymptotic mass of 
the quark in spacelike region  being the inverse of the  effective 
size of the pion. The SD equations also admit a chiral symmetric 
solution with confinement, this is understood to be an unstable 
vaccum. We also show that photon like behaviour of the gluon 
cannot cause chiral symmetry breaking for any value of the strong 
coupling constant $g$ in the rainbow graph approximation.

 Much of this analysis is done analytically by exploiting 
certain computational simplifications that we state in the
 paper. This simplification procedure can be applied for 
any relativistic field theory. Numerical investigation 
was essential to establish the existence of solutions 
in the space like region and thus estimate the asymptotic 
 masses of the quarks as a function of the parameters of the theory.
A recent review on SD equations is given in \cite{alkofer}.

\section{Lagrangian with String Tension}
The Minkowski Lagrangian of the theory is given by \cite{ramesha}

\begin{eqnarray}
Z&=&\int DA DQ DC D \overline \chi D \chi exp{(i S_0 +i \int j.( Q+A) )},
 \nonumber \\ 
S_0 &=& \int ({ -1 \over 4g^2} F ^2 + C_{\mu} (-D^2) Q^{\mu}
 + \overline \chi_{\mu}(- D^2) \chi^{\mu} - {\sigma \over 2} C_ \mu ^2 )  +\int ({\overline q} i\gamma_
\mu D^\mu  (A+Q)q - m \overline q q)
\end{eqnarray}
 where $F= F_{\mu \nu} ^a = \partial _\mu A_\nu ^a - \partial _\nu A_\mu ^a +f ^{a b c } A _\mu ^b A_\nu ^c$ 
is the antisymmetric gauge field tensor, $ A_\mu ^a$ are the 
gauge fields, $ g$ is the gauge coupling constant,
$f^{abc}$ are the structure constants of the corresponding 
non-abelian gauge group.  $Q_{\mu}$ and $C_{\mu}$ are bosonic
vector fields and $\overline{\chi}$ and $\chi$ are Grassmann 
Lorentz vector fields. All of them are in adjoint
representation of the gauge group and transform covariantly 
under local gauge transformations. The
significance of these fields is discussed in \cite{ramesha}. 
$D_\mu ^{a b} =D_{\mu}(A)=\partial _ \mu \delta ^ {a b} + f ^ {a c b
} A _ \mu ^c $ is the covariant derivative operator. $q$ and $\overline q$ 
are quark fields(fermions) in the
fundamental representation of the gauge group with a mass $m$.  
As can be seen from the Lagrangian the quark fields have a $
\overline q A q $ and a $ \overline q Q q $ interaction 
vertices.
$\sigma$ is the string tension which is asymptotically free \cite{ramesha}.
 Gauge fixing 
for the gluons has to be done just as in QCD and in the
Feynman gauge the $A_{\mu}$ and the $Q_{\mu}$ propagators are given by

 \begin{equation}
 < Q _\mu ^a Q _\nu ^b > = i\sigma { \eta _{\mu \nu} \delta ^{a b} \over {(q^2+i\epsilon)^2}} ~~\mbox{and}~~
< A _\mu ^a A _\nu ^b > = {-i \eta _ {\mu \nu} \delta ^{a b} \over {q^2+i\epsilon}}
\end{equation}
 
\begin{figure}[htbp]
\begin{center}
\epsfig{file=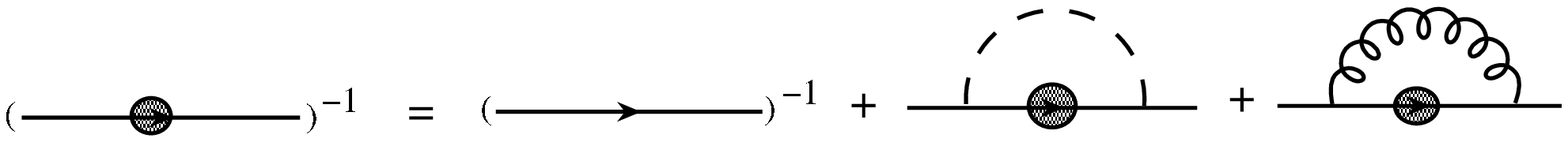,width=14cm,angle=0}
\end{center}
\caption{Schwinger Dyson Equation for Quark propagator}
\label{sdgraph}
\end{figure}

 We consider a systematic expansion in the various coupling constants 
$\sigma N$,$g^2 N$ and $1/N$ , where $\sigma$ is the
string tension, $g$ is the gauge coupling and $N$ is the number 
of colours, of which the leading
infrared(IR) contributions come from the $\sigma N$ term alone. 
Consequently we perform a non-perturbative expansion in $\sigma
N$ but perturbative expansion in $g^2 N$ and $1/N$ \cite{t'hooft}.  This yields 
for the quark propagator a summation of the rainbow graphs
with the $< Q _\mu ^a Q _\nu ^b >$ propagator alone . The 
SD equation can be considered with only this term alone
and solved
self-consistently. It turns out this equation has no UV 
infinite renormalisation.
But we know that including the $g^2 N$ term requires 
UV renormalisation.  This raises the question
whether non-perturbative IR physics is truly 
independent of the UV renormalisation procedure. 
To address this question in our analysis
we  include in the SD equation all the leading 
contributions in the various asymptotic IR and UV regions.  
The IR region is dominated by the $< Q _\mu ^a Q _\nu ^b >$ 
propagator whereas
the UV region is dominated by the $< A _\mu ^a A _\nu ^b >$ 
propagator  which gives the third term in the SD
equation.
It should be mentioned that to order $g^2 N$ in addition 
to the above term  there are some more graphs but
these are less important in the UV region and are not 
expected to make a significant contribution, for simplicity we neglect them.
In this work  we do not use running $\sigma$, $g$ or $m$ 
in our calculations. The SD equation
for the quark propagator with both the $< Q _\mu ^a Q _\nu ^b >$ 
and $< A _\mu ^a A _\nu ^b >$ propagator (Fig[\ref{sdgraph}]) is,

\begin{eqnarray}
S^{-1} (p)= \frac{1}{i}(\gamma .p - m) + \int {d^4k \over {(2\pi)^4}}({{\sigma N} \over {(k^2+i\epsilon)^2}}
\gamma_ \mu S(p-k) \gamma^\mu - {{g^2 N} \over {(k^2+i\epsilon)}} \gamma _ \mu S(p-k) \gamma^\mu)
\label{SD}
\end{eqnarray}
where $S(p)$ is the full quark propagator with momentum $p_{\mu}$ and mass $m$. 

The dashed lines in the figure corresponds to 
$<Q_{\mu}^a Q _\nu ^b >$ propagator and the springy line to the gluon.The 
solid line
with the blob is the full quark propagator and the 
solid line without the blob is the bare quark propagator.
Spin decomposing  the quark
propagator $S(p)$ as $ S(p)=i (A(p^2)\gamma.p +B(p^2)$) 
and substituting it in (\ref{SD}), we can express the SD  
equation as a coupled nonlinear equation in 
terms of two scalar functions $A(p^2)$ and $B(p^2)$.

\section{Causal Representation}
The solution space of SD  nonlinear equations  in general can 
be very large to address. Relativistic  Quantum Field Theory 
does impose certain qualifications and hence only those alone 
are of interest to us. Generally these restrictions are encoded 
in the Kallen-Lehmann representation. The physics consideration 
that go into this are that positive energies propagate forward 
in time and negative energies backward in time (as imposed by 
the Feynman $i \epsilon$ prescription) and all intermediate states have a 
positive norm. While these are essential for any physical propagators,
 it remains questionable whether unobservables should also have these 
properties. In particular the $\frac{1}{k^4}$ propagators necessarily
 generate negative norm intermediate states in perturbation theory \cite{dia}. 
Therefore we disregard the second assumption and  make the minimal 
assumptions necessary namely Lorentz covariance and Feynman $i \epsilon$ 
prescription yielding the causal representation.

\begin{eqnarray}
A(p^2) & = &\int_{-\infty}^{\infty} d\alpha \frac{\tilde{A}(\alpha)}{p^2-\alpha+i\epsilon}
\label{spec1}\\
B(p^2) &= &\int_{-\infty}^{\infty} d\alpha \frac{\tilde{B}(\alpha)}{p^2-\alpha+i\epsilon}
\label{spec2}
\end{eqnarray}
where $\tilde{A}(\alpha)$ and $\tilde{B}(\alpha)$ are   functions 
that are not required to be positive. Infact  the above are the 
Cauchy-Riemann integral representations for analytic functions. 
All we are demanding is that the quark propagator that we are 
interested in should be an analytic function of momentum. From 
the analyticity properties one can infer that the 
imaginary parts of $A(p^2)$ and $B(p^2)$ are the causal weight 
functions $\tilde{A}(\alpha)$ and $\tilde{B}(\alpha)$ themselves.
In eq.(\ref{spec1}) and  eq.(\ref{spec2}) out of generality we 
consider the entire range of $\alpha$. However as it turns out
 in the succeeding analysis $\tilde{A}(\alpha)$ and $\tilde{B}(\alpha)$ 
are non-vanishing only  for $\alpha >0$.

\section{ Regularisations}
The $<Q_{\mu}^a Q _\nu ^b >$ propagator behaviour $1/(k^2 +i\epsilon)^2$
is highly singular in the IR regime. In $3+1$    
dimensions the $+i\epsilon$ prescription is not sufficient to 
regulate it. Even after Wick rotation it has  a logarithmic singularity which 
makes this propagator ill-defined. Consequently the theory 
demands a regularisation of this IR divergence. There are many
equivalent ways of regularising this but we will mention the one
 which is convenient to implement in our  relativistic quantum 
field theory. In all loop calculations we are convoluting  this
 singular propagator with another quantity which has an   
integral representation which in its generic form is
 $\frac{1}{(k-p)^2-\alpha +i\epsilon}$ and higher powers thereof.
We illustrate the IR regularisation prescription by explicitly 
showing the steps in carrying out the $B(p^2)$ integration. 
Consider the following integral that occurs in the SD equation,
\begin{equation}
\int {d^4k \over {i (2\pi)^4}} \frac{B((p-k)^2)}{k^4}\\
\label{B}
\end{equation}
In eq.(\ref{B}), we substitute the  spectral representation 
eq.(\ref{spec2}) for $B((p-k)^2$ and using Feynman parametrisation,
 we  do the $1$-loop momentum integral explicitly by making the standard Wick rotation.
\begin{eqnarray}
\int_{-\infty}^{\infty}  d\alpha \int \frac{d^4k }{i (2 \pi)^4} \frac{1}{(k^2+i\epsilon)^2} \frac{\tilde 
B(\alpha)}{(k-p)^2-\alpha+i\epsilon}&=&\int_{-\infty}^{\infty}\frac{ d\alpha}{(4 \pi)^2} \int_{0}^{1} {dx
\over {(1-x)}} \hspace{0.2cm} \frac {x \tilde B(\alpha) }{(x p^2-\alpha +i\epsilon)}
\label{IR1}\\
&=&\frac{1}{(4 \pi)^2}\int_{0}^{1} {dx\over {(1-x)}} \hspace{0.2cm}  x \hspace{0.1cm} B(x p^2)
\label{IR2}\\
\nonumber
\end{eqnarray} 
In the above expression we have done the $\alpha$ integration 
formally to get back $B(p^2)$.
As is evident in eq.(\ref{IR2}) the IR singularity manifests
 itself as the end-point singularity at $x=1$ in the Feynman
 parameter $x$. The divergence is regulated by explicitly 
subtracting the $x=1$ singularity in the integrand.
\begin{eqnarray}
\int_{0}^{1} {dx\over {(1-x)}}   x  B(x p^2)&\rightarrow&\int_{0}^{1} {dx\over {(1-x)}}  (x  B(x p^2)-B(p^2))\\
\nonumber
&=&\int_{0}^{1} {dx \ln(1-x)} {d \over{dx}}(x B(x p^2))
\label{logrpn}\\
\nonumber
\end{eqnarray}
This defines the regularisation which can be  unambiguously 
implemented in the theory. The same procedure follows 
through  for the integration involving $A(p^2)$.

  The gluon term in the SD equation is UV divergent which 
needs to be regularised.  The standard dimensional regularisation
 and subsequent multiplicative renormalisation of the quantum fields 
and parameters can be implemented. We find the same can also be
 obtained in a uniform way by implementing the following representation for the gluon propagator

\begin{equation}
{1 \over {p^2+i\epsilon}} =  -\int_{0}^{\infty}d\beta \frac{1}{(p^2-\beta+i\epsilon)^2}
\label{glnrepn}
\end{equation}
Using the  causal representation we can do the loop  momentum integration.
The UV divergent terms are $\int_{0}^{\infty} d\beta A(\beta)$ and 
$\int_{0}^{\infty} d\beta B(\beta)$ which can be removed by 
renormalising the wave function and the quark mass parameter 
as in ordinary perturbation theory. The finite SD equation for  renormalised mass $m$ are,

\begin{eqnarray}
\frac{A(p^2)}{p^2 A(p^2)^2-B(p^2)^2} & = & 1+2 \bar{\sigma} \int_{0}^{1} dx \frac{x^2 A(xp^2)-A(p^2)}{(1-x)} - \bar{\alpha}
p^2 \int_{0}^{1} dx A(x p^2)(x^2-1)
 \label{ABfinal1}\\
\frac{B(p^2)}{p^2 A(p^2)^2-B(p^2)^2} & = & m+4\bar{\sigma}
\int_{0}^{1} dx \frac{x B(xp^2)-B(p^2)}{(1-x)} - 4 \bar{\alpha} p^2 \int_{0}^{1} dx B(x p^2)(x-1) \label{ABfinal2}
\end{eqnarray}
where  we have defined $\bar{\sigma}=\frac{\sigma N}{(4 \pi)^2}$ and $\bar{\alpha}=\frac{g^2 N}{(4\pi)^2}$ . In  the rest of the paper for convenience  
we will work in mass units of $\bar{\sigma}=1$.

\section{Confinement and Asymptotics}

All states which exist for asymptotic times are physical and
 they manifest as poles in the corresponding Green's functions.
 Physical quarks should be realised as poles in the quark propagators.
 Looking for poles in eq.(\ref{ABfinal1}), if $A(p^2)$ has a pole for 
some value of $p^2$, the left hand side of eq.(\ref{ABfinal1}) vanishes 
 and  the right hand side using the last representation in eq.(\ref{logrpn}),
 shows an edge singularity divergence. This contradiction in eq.(\ref{ABfinal1}) 
implies absence of pole like solutions. Similar is the case for $B(p^2)$. 
Finally if both $A(p^2)$ and $B(p^2)$ have poles at the same value of $p^2$ 
then by taking the ratio of eq.(\ref{ABfinal1}) and eq.(\ref{ABfinal2}) 
we note that again the equation cannot be matched. These observations 
make us conclude that $A(p^2)$and $B(p^2)$ cannot have poles, furthermore 
we conclude by refining this observation that $A(p^2)$ and $B(p^2)$  
cannot even have any divergent  behaviour for any value of $p^2$.

To analyse the  large space like $p^2$ ($p^2 \rightarrow -\infty$) 
behaviour of $A(p^2)$ and $B(p^2)$ in eq.(\ref{ABfinal1}) and 
eq.(\ref{ABfinal2}), we rescale the integrand $x \rightarrow \frac{x}{p^2}$ 
and obtain the approximate equation,
\begin{eqnarray}
\frac{1}{p^2 A(p^2)} & \sim & 1+2 \frac{1}{p^2} \int_{0}^{p^2} dx \frac{\frac{x^2}{p^4} A(x)-A(p^2)}{(1-\frac{x}{p^2})} - \bar{\alpha}
 \int_{0}^{p^2} dx A(x)(\frac{x^2}{p^4}-1) 
\label{asymA}\\
\frac{B(p^2)}{p^2 A(p^2)^2} & \sim & m+4 \frac{1}{p^2} \int_{0}^{p^2} dx \frac{\frac{x}{p^2} B(x)-B(p^2)}{(1-\frac{x}{p^2})} -4 \bar{\alpha}  \int_{0}^{p^2} dx B(x)(\frac{x}{p^2}-1) 
\label{asymB}\\
\nonumber
\end{eqnarray}
We further split the integral range to $(0,1)$ and $(1,p^2)$. 
It is easy to see that the latter dominates and we can  determine the 
asymptotic behaviour of $A(p^2)$ and $B(p^2)$ self-consistently 
in the UV regime. It is interesting to note that the UV behaviour of 
$A(p^2)$ and $B(p^2)$ is not influenced by the IR behaviour at 
all as expected in perturbation theory.

 For $p^2 \rightarrow -\infty$
\begin{eqnarray}
A(p^2) & \sim &\frac{1}{p^2}     ( 1,~ \frac{1}{ \sqrt{2\bar{\alpha} \ln(-p^2)}})
\label{asymbehA}\\
B(p^2) & \sim &  \frac{1}{p^2}     (m,~\ln(-p^2))
\label{asymbehB}\\
\nonumber
\end{eqnarray}
In the above equations the first term in the bracket refers to 
the asymptotic behaviour in the absence of the $\bar\alpha$ term 
while the second term refers to the asymptotic behaviour in the 
presence of the  $\bar\alpha$ term. It has to be noted that except 
in the chiral limit, upto logarithms the large momentum behaviour of 
$A(p^2)$ and $B(p^2)$ with and without the gluons are the same.

Next we look at the small momentum ($p^2 \rightarrow 0$) limit. 
From the integral equation the small $p^2$ behaviour of $A(p^2)$ 
and $B(p^2)$ are also self-consistently determined without any 
contribution from the UV region.  Infact  at $p^2 = 0$, $A(0)$ 
and $B(0)$ are determined to be 
\begin{eqnarray}\label{B0}
B(0)=\frac{m \pm \sqrt{m^2 +16}}{8}\\\nonumber
A(0)=\frac{B(0)}{m-B(0)}
\end{eqnarray}

Near $p^2=0$  we can have series expansion solution 
$A(p^2)=\sum_{n=0}^{\infty} a_{n} p^{2n} $ and $B(p^2)=\sum_{n=0}^{\infty}b_{n} p^{2n}$ 
for eq.(\ref{ABfinal1}) and eq.(\ref{ABfinal2}). Substituting the series and 
equating the coefficients of like powers of  $p^{2n}$, we get the following recursion 
relations for the coeffecients $b_{n}$ and $a_{n}$,

\begin{eqnarray}
a_{n}&=&
\frac{b_{n}+2 \sum_{m=0}^{n-1} (2 a_{m} b_{n-m} f_{n-m} - \bar{\alpha}(2 a_{m} b_{n-m-1} h_{n-m-1} - a_{n-m-1} b_{m} 
g_{n-m-1}))}{m- 4 b_{0}f_{0} +2 b_{0} f_{n+1}} \nonumber \\
& -& \frac{2\sum_{m=1}^{n}a_{n-m}b_{m} f_{n-m+1}} {m- 4 b_{0}f_{0} +
2 b_{0} f_{n+1}}
\label{smallA}\\
\nonumber
\end{eqnarray}

\begin{eqnarray}
b_n&=&
\frac{a_{n-1}-2 \sum_{m=0}^{n-1}( a_{m} a_{n-m-1} f_{n-m}+ 2\bar{\alpha} b_{m} b_{n-m-1} h_{n-m-1}) +2 
\bar{\alpha}\sum_{m=0}^{n-2} a_{m} a_{n-m-2} g_{n-m-2}}{m-4 b_{0} f_{0} -4 
b_{0} f_{n}} \nonumber \\
&+& \frac{ 
4\sum_{m=1}^{n-1} b_{m} b_{n-m} f_{n-m} } {m-4 b_{0} f_{0} -4 b_{0} f_{n}} 
\label{smallB}\\
\nonumber
\end{eqnarray}
where,
\begin{eqnarray}
f_i=\nonumber \sum_{t=0}^{i} {1 \over t+1};h_i={1 \over (i+1)(i+2)}; g_i=\nonumber{1 \over (i+1)(i+3)}\\
\nonumber
\end{eqnarray}
 The radius of convergence of the  series solution is difficult to 
ascertain analytically. Investigating it numerically, we find that 
 for small masses $m$, the series converges in a small interval 
around $p^2=0$. As we increase the mass $m$, the interval 
of convergence also increases. We attribute the finite radius of 
convergence to the onset of the  imaginary part  in  the SD 
equation at some value of the $p^2>0$ and this value is  
proportional to the mass $m$. The onset of this behaviour is
 perhaps  due to a branch cut and consequently no Taylor expansion
 can  converge.  Numerical solution of the integral equations 
eq.(\ref{ABfinal1}) and eq.(\ref{ABfinal2}) support this assumption.
 The onset of non-analyticity in $A(p^2)$ and $B(p^2)$ in general can
 be different but we find that for small quark mass $m$ they are 
about the same. We cannot ascertain precisely the nature of  
non-analyticity but it can be $A(p^2) \sim const + \frac{1}{ln(\tilde{m}^2-p^2)}$ 
or even softer. This follows because if we assume that $A(p^2)$ or $B(p^2)$ 
is singular at some value of $p^2$, the $r.h.s$ of eq.(\ref{ABfinal1}) and 
eq.(\ref{ABfinal2}) will yield even more singular contribution for the 
same value of $p^2$ thus mismatching the equation. Similar is the case for $B(p^2)$. 
The Taylor expansion suggests that for small quark masses
$m$, for  both $A(p^2)$ and $B(p^2)$ functions the non-analyticity
threshold is at $\tilde{m}^2$ and is numerically estimated to be
\begin{eqnarray}
\tilde{m}^2 & \sim  &.02+ m * const\hspace{0.5cm} for  \hspace{0.5cm} m \ll 1  \\
\nonumber
\tilde{m}^2 & \sim & m^2 \hspace{0.5cm} for \hspace{0.5cm}  m>1
\nonumber
\end{eqnarray}
For large quark masses we estimate  $\tilde{m}^2$ by doing the naive 
large $m$ or equivalently the small $\sigma$ and $\alpha$ perturbation
 theory. In this perturbation theory where $\tilde \sigma$ is also small ,
 it is easy to recognise that $A(p^2)$ and $B(p^2)$ may not have strict 
poles but poles softened by logarithms such as $\frac{\ln{(p^2-m^2)}}{p^2-m^2}$. 
Hence there is a net singularity softer than a pole. This is a remarkably 
different behaviour as compared to that of light quarks. 
 
In eq.(\ref{B0}), if we pick $B(0)$ negative solution, we are able to 
establish analytically that the imaginary part may exist only for $p^2>0$ 
and both $A(p^2)$ and $B(p^2)$ are negative for space like $p^2$ . 
From these two observations it is self-consistent to infer that  
that the imaginary part of $A(p^2)$ and $B(p^2)$ are also positive.
 Consequently we can conclude that the quark propagator in our theory
 is consistent with the standard Kallen-Lehmann representation.

\section{Numerical Solutions} 

The numerical solutions for $A(p^2)$ and $B(p^2)$ in the 
space-like($p^2 < 0$) region can be
obtained by iteration. Solutions exist with or without 
the gluon term. We ignore the gluon term for simplicity .
The functions $A(p^2)$ and $B(p^2)$ are recast in terms of 
two new functions  $\bar A(p^2)$ and 
$\bar B(p^2)$ through the following transformation

\begin{eqnarray}\label{newAB}
A(p^2)=\frac{\bar A(p^2)}{p^2 \bar A(p^2)^2- \bar B(p^2)^2}\nonumber\\
B(p^2)=\frac{\bar B(p^2)}{p^2 \bar A(p^2)^2- \bar B(p^2)^2}
\end{eqnarray}
This is substituted in eq.(\ref{ABfinal1}) and solved for 
$\bar A(p^2)$ and $\bar B(p^2)$ by giving a
seed which is $\bar A(0)$ and $\bar B(0)$. Such an  equation 
is numerically convenient to iterate. $A(p^2)$ and 
$B(p^2)$ can be obtained by solving eq.(\ref{newAB}) in 
terms of $\bar A(p^2)$ and $\bar B(p^2)$. The 
nature of the integral equation(\ref{ABfinal1}) is 
such that to solve it at a particular value of momentum,  
one has to know the functions at all values of
momentum below that particular value all the way upto $p^2=0$. The 
momentum $p^2$ is discretised and the
integration over $x$ is done numerically. We use an interpolation 
formula to fit the iterate and do the
$x$ integration. For values of $m>0.5$ the iteration method works very well. 
But for small values of the
quark mass $m$, due to the onset of the imaginary part very 
close to $p^2 =0$, one has to be very careful. The solutions for
 very light masses have been determined with about  $5\%$ inaccuracy.

\begin{figure}[hH]
\parbox{0.19\textwidth}
  {\epsfig{file=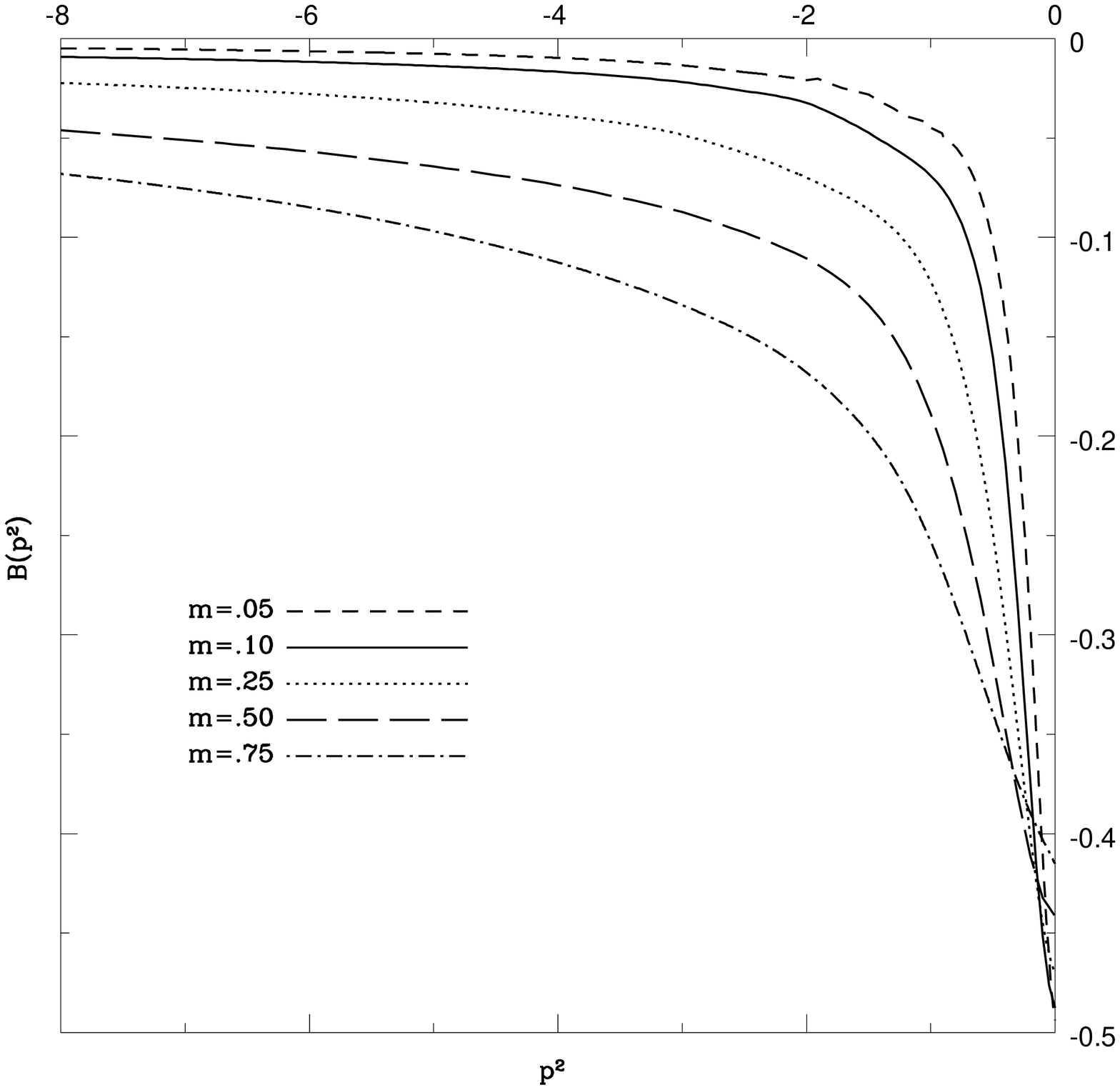,width=0.49\textwidth}}\hfill
\parbox{0.49\textwidth}
  {\epsfig{file=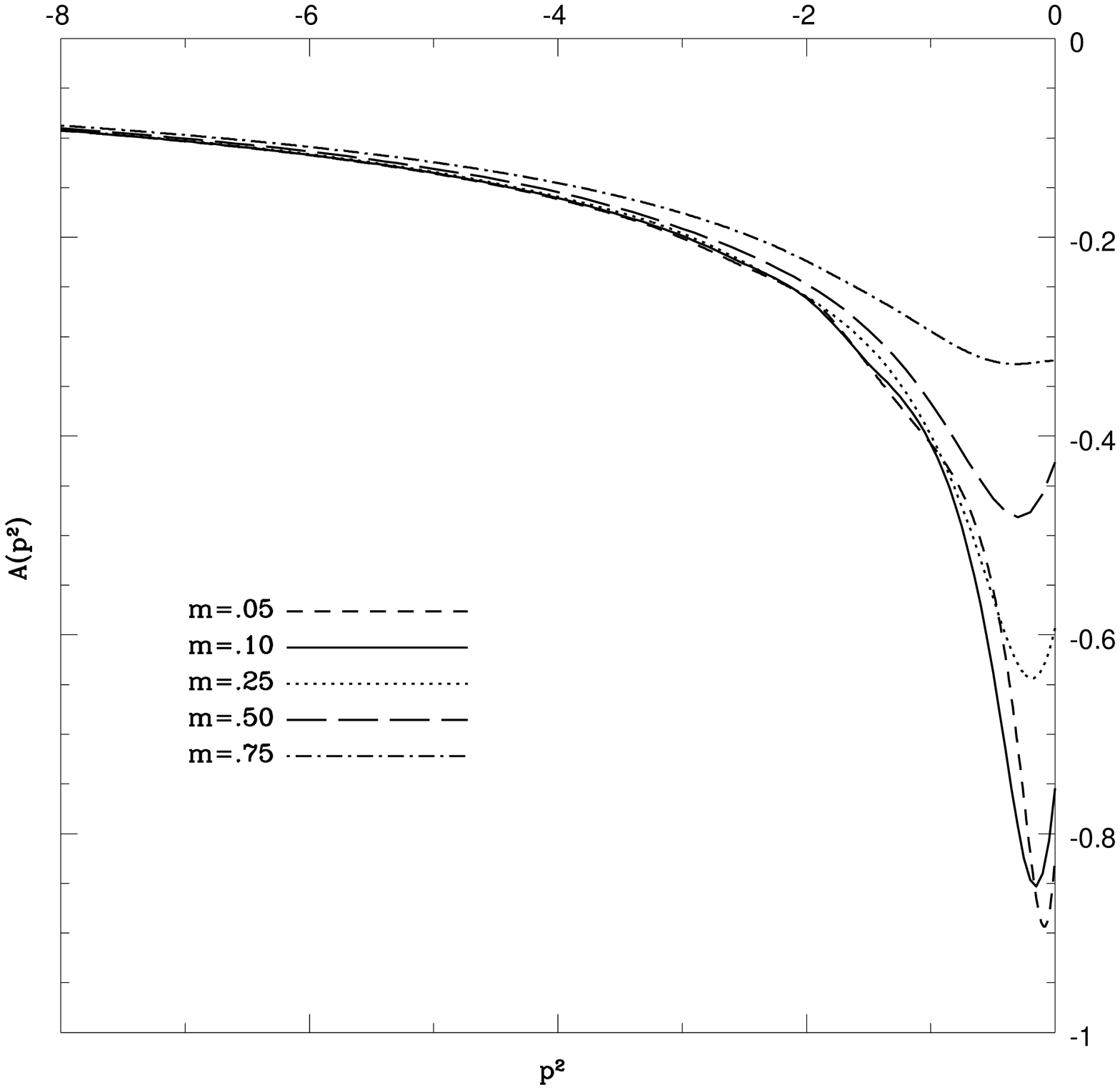,width=0.49\textwidth}}

\parbox{.5\textwidth}{\raisebox{1ex}[0ex][0ex]{\large (a)}} 
\parbox{.3\textwidth}{\raisebox{1ex}[0ex][0ex]{\large (b)}}

\caption{ Numerical solutions for (a) $B(p^2)$ and (b) $A(p^2)$ for light masses .}
\label{light}
\end{figure}

The solution for light quark masses are shown in Fig[\ref{light}a] and Fig[\ref{light}b].
 The string tension $\bar \sigma$ sets the scale for small $p^2$ 
and the dominant contribution in this region comes from this term. 
For $p^2>1$ the momentum $p^2$ sets the scale. The large momentum 
behaviour given by eq.(\ref{asymbehA}) and eq.(\ref{asymbehB}) 
sets in at around $p^2\sim 3$.
For small values of $m$, $A(p^2)$  has a minima. For masses 
$m>0.82$ this minimum does not exist in the spacelike region.

\begin{figure}[hH]
\parbox{0.19\textwidth}
  {\epsfig{file=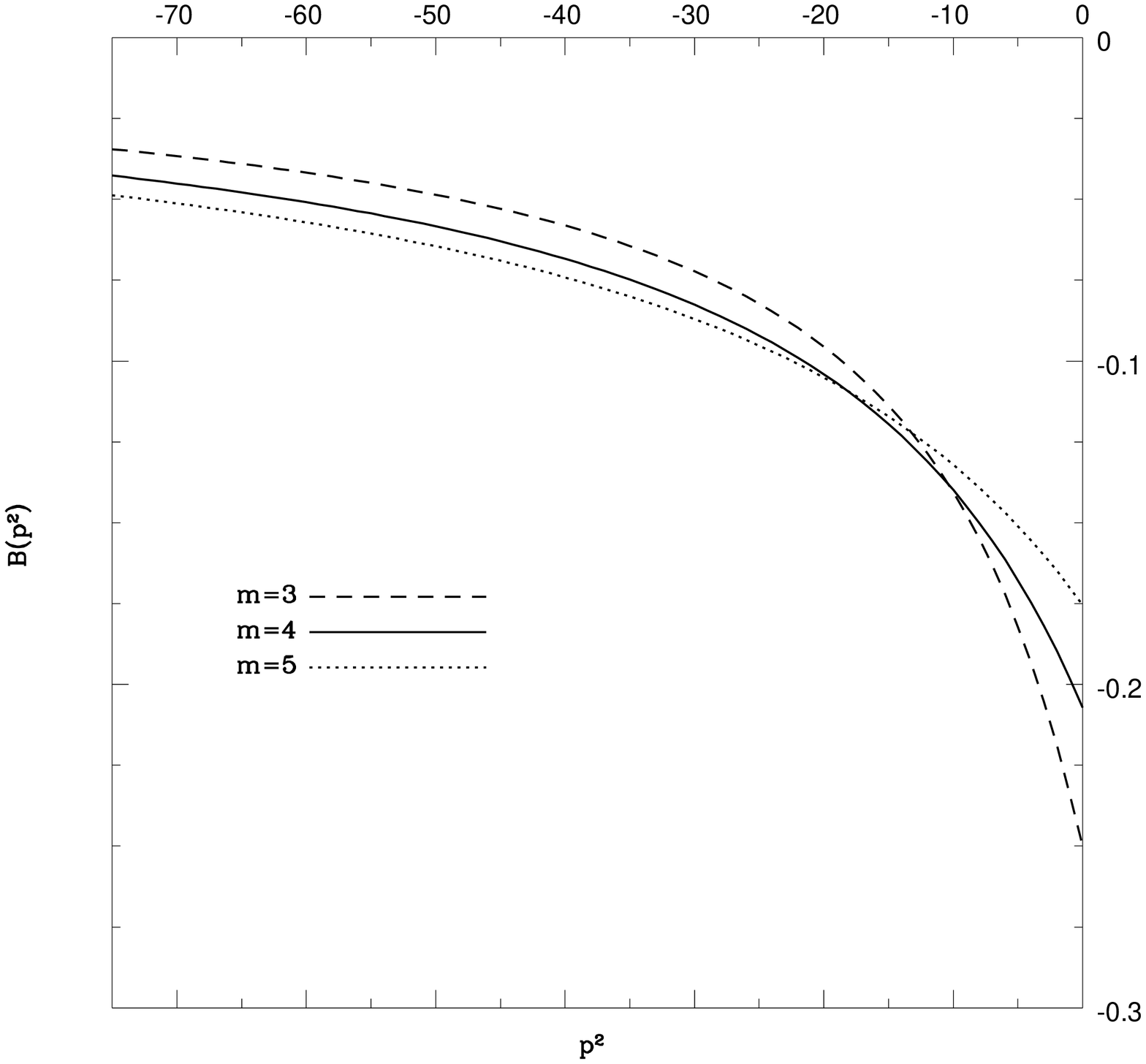,width=0.49\textwidth}}\hfill
\parbox{0.49\textwidth}
  {\epsfig{file=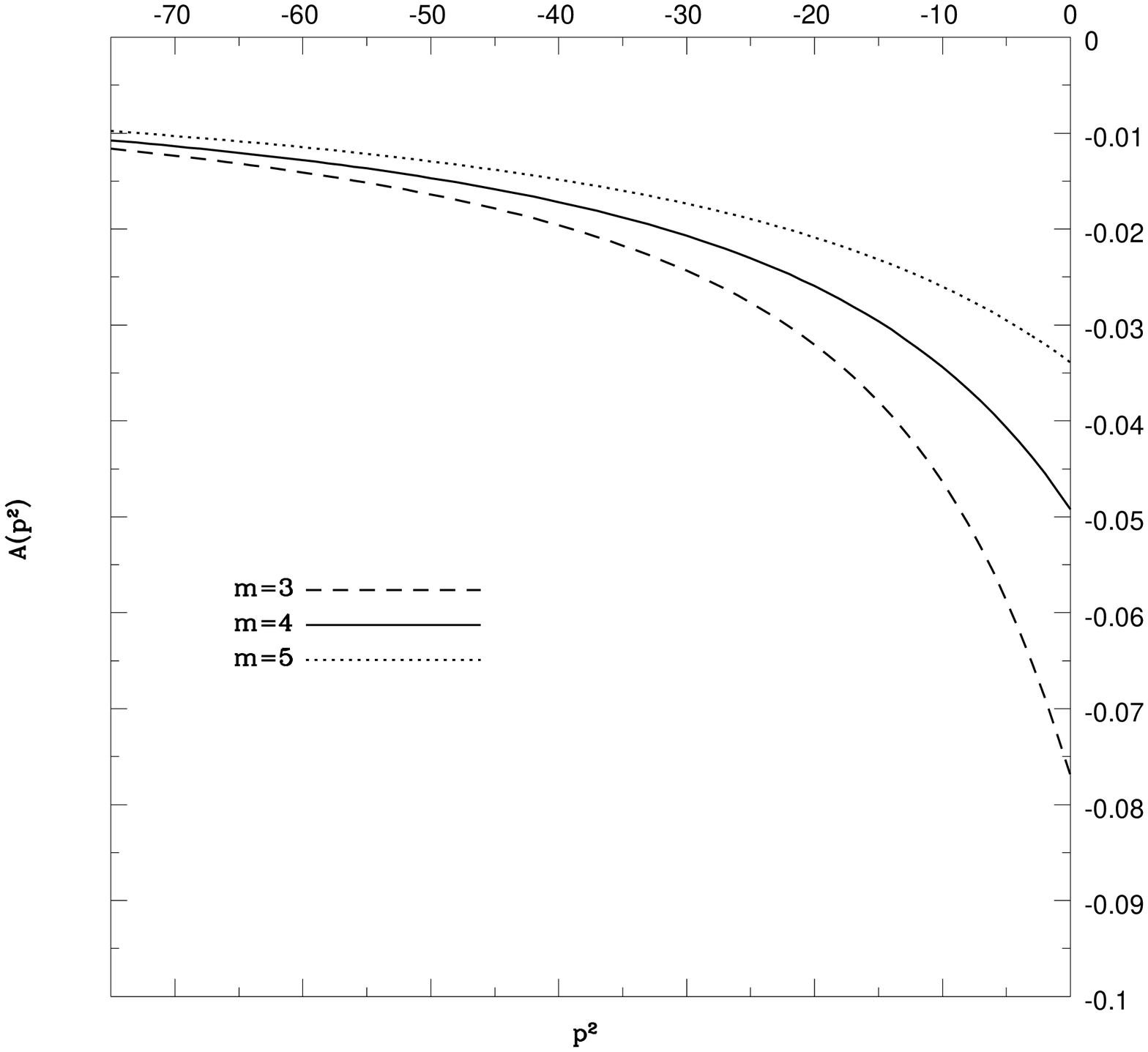,width=0.49\textwidth}}

\parbox{.5\textwidth}{\raisebox{1ex}[0ex][0ex]{\large (a)}} 
\parbox{.3\textwidth}{\raisebox{1ex}[0ex][0ex]{\large (b)}}

\caption{ Numerical solutions for (a) $B(p^2)$ and (b) $A(p^2)$ for Heavy masses .}
\label{heavy}
\end{figure}

The solutions for heavy quark masses, are shown in Fig[\ref{heavy}a] and Fig[\ref{heavy}b]. 
Unlike in the case of light 
quarks,  the mass $m$  dominates over the string tension and  sets 
the scale for small value of momentum. For large momentum like in 
the light quark case, its the momentum that sets the scale. 

It can be noted that  all the solutions for $A(p^2)$ and $B(p^2)$ 
in the large momentum limit goes over smoothly to the 
asymptotic expressions in (\ref{asymbehA}) and(\ref{asymbehB}). 
Upto to logarithmic corrections, $A(p^2)$ and $B(p^2)$  have  
essentially the canonical $\frac{1}{p^2-M^2}$  behaviour in the 
space-like region for all masses $m$ where $M$ is called the 
asymptotic mass. A useful parametrisation for $A(p^2)$ and $B(p^2)$ is,
\begin{eqnarray}\label{effans}
A(p^2)\sim \frac{-A(0) M_{a}^2}{p^2-M_{a}^2}\nonumber\\
B(p^2)\sim \frac{-B(0) M_{b}^2}{p^2-M_{b}^2}
\end{eqnarray} 

\begin{figure}[hH]
\begin{center}
\epsfig{file=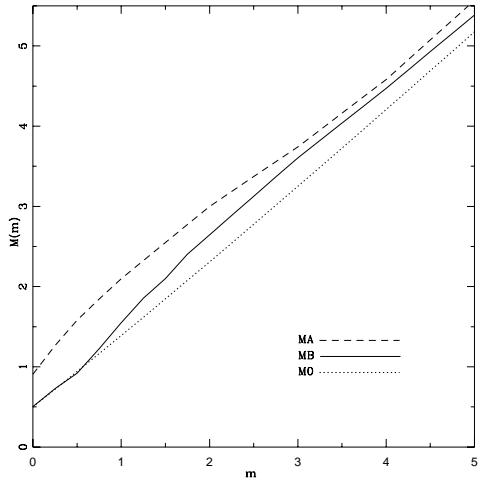,width=7cm,angle=0}
\end{center}
\caption{Asymptotic mass  $M_a$, $M_b$,$M_0$ for $A(p^2)$ and $B(p^2)$ $v/s$ renormalised mass }
\label{effmass}
\end{figure}

The ``asymptotic masses" $M_{a}^2$ and $M_{b}^2$  depend on the 
mass $m$. A convenient way of determining $M_a^2$ and $M_b^2 $ is by 
defining them to be that value of $p^2$  where $A(p^2)$ and $B(p^2)$ 
are reduced by half from their values at $p^2=0$. Doing so we can infer 
the asymptotic masses $M_a$ and $M_b$ from the numerical data. In Fig[\ref{effmass}] 
we have plotted $M_a$ and $M_b$ versus $m$ and  for comparison 
we have also plotted $M_0=\frac{B(0)}{A(0)}$ which is 

\begin{eqnarray}
M_0=\frac{7 m + \sqrt{ m^2+16}}{8}
\end{eqnarray}
 The asymptotic  mass does increase almost linearly with $m$, all 
through from light to heavy quarks. There are two asymptotic  masses 
for any quark, namely $M_{a}$, the spin-dependent asymptotic  mass 
and $M_{b}$ the spin-independent asymptotic  mass and the former is  
significantly heavier than the latter for light quarks. 

We can extend the numerical ansatz to small $p^2>0$ as well. Beyond 
which as we understand from the Taylor expansion the numerical 
solutions cannot converge . At this point we have to work with 
the real and imaginary parts of $A(p^2)$ and $B(p^2)$ which is 
numerically cumbersome to handle. The asymptotic  mass behaviour 
of the quark propagator given in eq.(\ref{effans}) is to be taken 
for spacelike $p^2$ only. For timelike $p^2>0$ this is qualitatively unacceptable due to confinement.

 The above procedure of determining numerical solutions can be 
done for $B(0)$ positive also. We find that the numerical solution 
is not so stable. This is partly because in the the $r.h.s$ of 
eq.(\ref{ABfinal2}), we can see that the integral contribution 
is of opposite sign to that of the constant term $m$. Therefore 
at some point $B(p^2)$ will vanish and change sign. The existence
 of this solution can be doubted numerically but it cannot be 
ascertained. In any case this solution is not of physical interest 
as it disobeys Kallen-Lehmann positivity condition. From eq.(\ref{spec1})
 and eq.(\ref{spec2}), knowing that $\tilde A(\alpha)$ and $\tilde B(\alpha)$ 
are non vanishing for $p^2> \tilde{m}^2>0 $,  $A(p^2)$ and $B(p^2)$ 
will never vanish in the space like region if $\tilde A(\alpha)$ and $\tilde B(\alpha)$ are positive.

\section{Chiral Limit}

We look at the theory when the bare mass of the quark vanishes. 
Then the theory has a global continuos symmetry namely chiral 
symmetry. In our large $N$ and small $g^2 N$ expansion, doing 
just the wave function renormalisation, we obtain the SD equations
 where we have  eq.(\ref{ABfinal1})  and
\begin{eqnarray}
\frac{B(p^2)}{p^2 A(p^2)^2-B(p^2)^2} & = & 4 \bar{\sigma}\int_{0}^{1} dx \frac{x B(xp^2)-B(p^2)}{(1-x)} + 4 \bar{\alpha} \int_{0}^{1}dx \int_{-\infty}^{x p^2}d\beta B(\beta)
\label{chiralB}\\
\nonumber
\end{eqnarray}

The last term has  a constant perhaps finite or infinite. 
If it is infinite, then renormalisation is necessarily demanded.
 If it is finite renormalisation is an option. In either cases 
if we do any renormalisation, finite or infinite, it is evident 
that this implies we have formally included a bare mass term and 
then subsequently arranged the renormalised mass to vanish. This 
procedure cannot respect chiral symmetry.  This is easy to see by 
noting that the massless Goldstone meson wave function satisfies 
the following Bethe-Salpeter equation \cite{BS},
\begin{equation}
\phi(p^2)=(p^2 A(p^2)^2-B(p^2)^2)(4 \bar{\sigma}\int_{0}^{1} dx \frac{x \phi(xp^2)-\phi(p^2)}{(1-x)} + 4 \bar{\alpha} \int_{0}^{1}dx \int_{-\infty}^{x p^2}d\beta \phi(\beta))
\label{pion}
\end{equation}
 In a renormalisable theory, $\phi(p^2)$ does not undergo any other 
renormalisation. Consequently in eq.(\ref{pion}) the last term 
cannot be altered. It is clear that the solution to eq.(\ref{pion}) 
is $\phi(p^2)=B(p^2)$ provided we do not make any finite or infinite 
renormalisation of eq.(\ref{chiralB}). $\phi(p^2)=B(p^2)$ is precisely 
the consequence of the  Goldstone theorem. This property of SD and Bethe-Salpeter
equations, first realised by
Mandelstam \cite{mand2},  is true 
in theories with spontaneous symmetry breaking in the rainbow graph
approximation. To reiterate, any mass 
renormalisation scheme if adopted such that the constant term 
in the  last term in eq.(\ref{chiralB}) is effectively absent 
then chiral symmetry is explicitly broken \cite{pagels,lane}.

 Alternatively we can investigate if eq.(\ref{chiralB}) as such 
has any solutions. The crucial  property for the existence of 
the solution is its UV behaviour. Using  exactly the same  
approximation as in eq.(\ref{asymA}) and eq.(\ref{asymB}) 
we find that behaviour of $A(p^2)$ for large spacelike 
momentum is unchanged from eq.(\ref{asymbehA}), while $B(p^2)$ 
behaves as 

\begin{eqnarray}
B(p^2) & \sim &  \frac{const}{(p^2)^2}     (\frac{1}{p^2},\frac{1}{(\ln(-p^2))^3})
\label{chiralasym}
\end{eqnarray}
The UV behaviour of  $B(p^2)$ is indeed more convergent 
than the canonical behaviour. Consequently the last term 
in eq.(\ref{chiralB}) is finite. Thus our system is self 
consistently determined with only wave function renormalisation 
and no mass renormalisation.

 In the small $p^2$ regime not much of a difference occurs and 
indeed $B(0)=\pm\frac{1}{2}$ ( neglecting the $\bar \alpha$ term) 
as expected from eq.({\ref{B0}). Asymptotic behaviour of $B(p^2)$ 
as in eq.(\ref{chiralasym}) is determined upto a constant and this 
can have either sign. Now it is possible to interpolate both the 
solutions to their corresponding asymptotic behaviour without 
$B(p^2)$ vanishing. These solutions are chiral conjugate. 
This is because  in eq.(\ref{chiralB}) there is an ambiguity in 
the sign of $B(p^2)$. Both $B(p^2)$ and $-B(p^2)$ are possible. 
Thus picking one solution breaks the chiral symmetry spontaneously. 
Note
that $A(p^2)$ remains the same irrespective of the choice of
$B(p^2)$, because by definition $A(p^2)$ is chirally symmetric.
Both $A(p^2)$ and $B(p^2)$  have analytic expansions around 
$p^2=0$ as determined from the series solution, eq.(\ref{smallA}) 
and  eq.(\ref{smallB}) with $\bar \alpha=0$. The radius of 
convergence of this series is numerically inferred to be 
nonvanishing as shown in eq.(\ref{smallB}) for $\bar{\alpha}=0$. 

        Since the asymptotic behaviour of $B(p^2)$ 
is very different from the canonical, we should 
expect the asymptotic mass definition to be 
modified. Namely for space like $p^2$
 
\begin{eqnarray}
B(p^2) & \sim &  \frac{B(0)(M_{b}^2)^2}{(p^{2}-M_{b}^2)^2}     
(\frac{-M_{b}^2}{(p^2 -M_{b}^2)},~\frac{1}{(\ln(-p^2))^3})
\end{eqnarray}
Numerically the asymptotic mass can be estimated 
and it is found that $M_{b}^2 \sim (0.7- 0.9)$.
The qualitatively different behaviour of $B(p^2)$ in 
the exact chiral limit cannot be obtained continuosly by 
taking the renormalised mass $m\rightarrow 0$. Hence this 
theory exhibits critical behaviour at $m=0$. This as we 
see is a consequence coming totally due to the absence 
of mass renormalisation in contrast to the massive case. 
The chiral limit exhibits spontaneous symmetry breaking 
but also shows that the spin  independent propagation 
of a quark is non-analytic as a function of the quark 
mass $m$ at $m=0$. Inspite of this in the IR region as 
seen from the Taylor series we do not see any signature 
of the critical behaviour.

  SD equations (\ref{ABfinal1}) and (\ref{chiralB}) also 
have chiral symmetric solutions wherein $B(p^2)$ vanishes 
for all $p^2$ and $A(p^2)$ is non-vanishing. We can show there 
exists a series solution for 
$A(p^2)=\frac{1}{\sqrt{-p^2}} \sum_{n=0}^{\infty}a_n (-p^2)^\frac{n}{2}$ 
for small $p^2$ and the asymptotic behaviour remains unchanged 
as in (\ref{asymbehA}). This solution preserves chiral symmetry. 
Note that this is also a confining solution as there is no pole 
for the quark propagator. Indeed the propagator is finite at 
$p^2=0$ and has only a square root branch cut starting from 
the origin $i.e.$, the threshold mass vanishes. SD equations 
can have many solutions. Each of them actually correspond to 
different possible stable or unstable vacua. The criterion 
to pick the minimum vaccum configuration cannot be inferred 
from the SD equations alone. This can be carried out by 
summing the vaccum graphs which in turn depend on the 
propagator functions. We have not attempted this analysis 
here. By our expectations either we have one solution to 
the SD equation in which case that will correspond to the 
vaccum or there are three (odd) solutions of which  two are 
related by chiral symmetry and hence to be degenerate stable 
vacua and the third has to be unstable. Any choice of the 
stable vacua  therefore will yield symmetry breakdown and 
the chirally symmetric choice is unstable.

We now address symmetry breakdown between two stable vacua 
in the absence of string tension. In eq.(\ref{chiralB}) we 
put $\bar \sigma =0$. This is a  truncation of QCD in which  
we are keeping the rainbow graphs due to gluons alone. This 
is not a controlled approximation but it is interesting to 
know whether less singular interactions with sufficiently 
large coupling constant $\bar \alpha$ can cause spontaneous 
symmetry breakdown. The answer as we prove now is that it 
cannot cause symmetry breakdown. The asymptotic UV behaviour 
remains the same as before. In the absence of the 
$\bar \sigma $ term we do expect the quark propagator to obey 
the Kallen-Lehmann representation, consequently for space like 
$p^2$, $A(p^2)$ and $B(p^2)$ cannot have any other zeroes other 
than at asymptotic limit. Furthermore we also know if $B(0)$ is 
positive(negative) and it remains positive(negative) in the 
entire space like regime, due to Kallen-Lehmann representation. 
The consequence in eq.(\ref{chiralB}) at $p^2=0$ is that the 
$r.h.s$ is negative(positive) but the $l.h.s$ is positive(negative) 
showing an inconsistency. Thus the only consistent alternative is 
$B(0)=0$ and hence no chiral symmetry breakdown.

\section{Discussion}
 We speculate some consequences of our semi-analytic analysis 
supported by numerical estimates for strong interaction 
phenomenology. In \cite{BS} by fitting the vector meson 
mass we estimate $\tilde m \sim .5 Gev$. This threshold 
mass is not a physical mass accessible at asymptotic times, 
however in most dynamics of quarks it still should give 
similar effects in the transient existence. For example 
in stable bound states quarks would have to have energy 
less than $\tilde m$. Consequently an electromagnetic 
probe will find the quark degrees of freedom to 
have a ``constituent'' mass depending upon  the 
bound state but necessarily less than the threshold 
mass. Similarly    a high energy quark will radiate 
and  lose energy but  it will cease to radiate when 
it reaches the energy of the threshold mass.

 From  \cite{BS} the light quark ($u$ and $d$) renormalised 
mass is estimated from the pion mass to be about $6Mev$. This 
is consistent with other phenomenological considerations \cite{leut} 
and is substantially different from $\tilde m$. In this theory 
string tension runs and it decreases at high energies. 
  We define 
heavy quark mass in comparison to relevant string tension. If $m^2> \bar \sigma$, 
we find that all notions of masses $m$, $\tilde m$ and $M$  are the same. 
Furthermore heavy quarks in some sense are not severely effected 
by string tension. Apart from  confinement, it is suggested that heavy 
quarks have a longer transient existence in contrast to  light quarks.

 String tension, naively can cause non-unitary behaviour 
in the theory. We find that with in this limited exercise 
of quark propagator and their bound states many of the 
consequences of unitarity still hold true although we 
do not have any formal proof. This issue should be 
understood more carefully.  Infact  we find numerically 
the solution of the SD equation which is consistent 
with the requirements of unitarity is unique and stable. 
While the other possible solutions are numerically unstable.
 
 Chiral limit of the theory can be consistently defined 
only if there is no mass UV renormalisation. For this 
reason alone chiral limit is a critical point since 
for any other massive case the theory is defined 
only by mass renormalisation. Inspite of this the 
infrared behaviour of the quark propagator smoothly 
extrapolates from small mass to vanishing mass. 
Because of this PCAC(Partial Conservation of Axial 
Current) relations still hold. We find chiral 
symmetric solution of the SD equation and it is 
also confining due to the absence of a pole but it
  is expected to be an unstable vaccum.

 The techniques of solving SD equations that we 
have enunciated here is fairly general. The massless 
particle exchange that is considered explicitly has 
some algebraic simplifications. Essentially eq.(\ref{IR1}) 
and the trick of eq.({\ref{glnrepn}) can be adopted for 
any generic relativistic field theory to get equations 
of the type (\ref{ABfinal1}) and (\ref{ABfinal2}).

   All our considerations above are equally applicable 
to techincolour models \cite{Susskind}. Indeed the existence of many 
mass scales $m$, $\tilde m$, and $M$ is the scenario 
suggested in the literature, here we have a concrete 
compatible model where it is realised.

\end{document}